\documentclass{kluwer}
\usepackage[dvips]{graphicx}
\newdisplay{guess}{Conjecture}

\def\beq{\begin{equation}}
\def\enq{\end{equation}}
\def\beqn{\begin{eqnarray}}
\def\enqn{\end{eqnarray}}
\def\eps{\varepsilon}

\begin{document}

\begin{opening}

\title{X-ray Lines in Gamma-ray Bursts and Cerenkov Line Mechanism}
\author{Wei Wang}
\institute{National Astronomical Observatories, Chinese Academy of Sciences, Beijing 100012, China  \\
Department of Physics, University of Hong Kong, Pokfulam Road, Hong Kong \\
wwang@bohr.physics.hku.hk}
\date{May 20, 2004}

\begin{abstract}
X-ray emission and absorption features are of great importance in
our understanding the nature and environment of gamma-ray bursts
(GRBs). So far, iron emission lines have been detected in at least
four GRB afterglows. In this paper, the observational properties
and physical constraints on materials surrounding GRB sources are
reviewed, and several classes of theoretical models are also
discussed. We will specially concentrate on the Cerenkov line
mechanism, in which the broad iron lines are expected, and a small
mass of Fe is required to produce the large line luminosity. In
addition, our interpretation can favor the recent jet unified
model for different classes of gamma-ray bursts with a standard
energy reservoir.
\end{abstract}


\end{opening}

\section{X-ray lines in gamma-ray bursts}
The detection of X-ray emission lines and absorption features in
the GRB afterglows provides a powerful tool to probe the
environments around GRB sources, and then probably gives us clues
to the nature of the progenitors. Moreover, the identification of
X-ray lines by the high spectral resolution detectors could
directly provide the redshift information for GRBs.

Recent GRB observational reports have shown the emission line
features in the X-ray afterglows of at least four GRBs (GRB
970508, Piro et al. 1999; GRB 970828, Yoshida et al. 1999; GRB
991216, Piro et al. 2000; GRB 000214, Antonelli et al. 2000).
These lines generally last half day to several days, with large
line width and very high line luminosity. The properties of the
detected emission lines in GRBs are summarized in Table 1. The
X-ray prompt emission is also detected in GRB 990712 with BeppoSAX
(Frontera et al. 2001), which could be interpreted as either a
line profile with centroid energy of 4.5 keV or a blackbody
spectrum with the temperature $\sim 1.3$ keV, so we will not
discuss it further in this paper. According to the variability and
very large flux and equivalent width, we can estimate that there
are large iron masses of 0.01-1.0$M_\odot$ in the emission region
with a size of $\sim 10^{15-16}$ cm if these lines are supposed to
be produced by fluorescence or recombination. So much Fe would
come from the surrounding medium rather than GRB progenitors
themselves, because it will involve the famous problem of baryon
contamination.

A large number of irons in the close environment of GRBs imply the
mental enriched process before GRBs occur. Then there exists a
very dense medium surrounding the GRB progenitor, so we expect
that the absorption features should be detected too. Up to now,
only a discovery of a transient absorption edge in X-ray spectrum
of GRB 990705 (Amati et al. 2000) has been reported. The
absorption feature shows that the iron abundance is about 75 times
higher than the solar value. Other ion emission lines were also
detected in GRB 011211 by an XMM-Newton observation (Reeves et al.
2002). The X-ray spectrum reveals evidence for emission lines of
Mg, Si, S, Ar, Ca and possible Ni. The line identification show
the redshift $z=1.88\pm 0.06$, differing significantly from the
known GRB redshift $z=2.140\pm 0.001$ by the spectroscopy of the
optical afterglow (Holland et al. 2002), implying an outflow
velocity for the line emitting material of $v = 25800{\rm km\
s^{-1}}$ or $v/c=0.086$. Additional evidence for the presence of
ion lines were also suggested for two XMM-Newton spectra of GRB
001025A and GRB 010220 (Watson et al. 2002). These ion lines are
attributed to the emission from the ionized optically thin plasma,
indicating that thermal emission from light elements may be common
in the early X-ray afterglows. And metal enrichments may occur
preceding the GRB events.

\section{Model constraints from observations}
The iron line observations show constraints on the theoretical
models: (1) If the iron lines are produced by fluorescence or
recombination, we require that the compact regions ($\sim
10^{15-16}$ cm) contain $0.01-1 M_\odot$ Fe with the electron
density larger than $10^{10}{\rm cm^{-2}}$, but must be optically
thin to electron scattering (size problem). (2) So much iron
(corresponding to several solar mass material) cannot be produced
by burst itself because of the famous problem of baryon
contamination. These irons should come from the surrounding
medium. Furthermore, iron-rich medium may be not consistent with
the normal interstellar medium even in the dense region of stellar
formation. (3) If we interpret the line width as due to the
velocity of a supernova remnant, their sizes allow us to estimate
the age of SNR which turn out to be tens of days, while at this
time, cobalt nuclei would dominate and the line would be produced
mainly by cobalt not iron (kinematic problem).

A large mass of Fe around the progenitor really produces a
challenge to the present GRB models. And the problem also focuses
on how the torus of iron-rich material formed. M\'esz\'aros \&
Rees (1998) have shown that the circumburst environment created by
the stellar wind before the hypernova (Paczy\'nski 1998). While,
this scenario may not produce so high luminosity Fe lines.
Recently, Dado, Dar \& De R\'ujula (2003) argued that the line
detected are not the real iron lines but the strongly
Doppler-blueshift Ly$\alpha$ line emissions by jets of highly
relativistic cannonballs produced by a supernova. At present,
there are mainly two classes of model put forward to explain the
lines: the geometry dominated and engine dominated models, and the
key difference is the reason for the observed duration of the line
emission.

In the geometry dominated models, the line emission duration is
due to light travel time effect. Since the lines appear in the
timescale of one day, the emission size estimated is about
$10^{16}$cm. A favorable situation is expected in the supranova
(Vietri \& Stella 1998), where a GRB is preceded by a supernova
explosion for several months to years with ejection of an
iron-rich massive shell. In this scenario, high density in excess
of $10^{10}{\rm cm^{-3}}$ may be expected, and Fe absorption
features of strongth comparable to these in emission should be
also be detected. While, in the engine dominated models, the
emission material, at a typical emission size about $10^{13}$cm,
can be pre-GRB ejecta or the expanding envelope of a supergiant
GRB progenitor (Rees \& M\'esz\'aros 2000, M\'esz\'aros \& Rees
2001). They have argued that the strong line emission can be
attributed to the interaction of a continuing post-burst
relativistic outflow from the central decaying magnatar with the
progenitor stellar envelope at distance less a light hour, and
only a small mass Fe is required.

\begin{figure}
\centerline{\includegraphics[width=7cm]{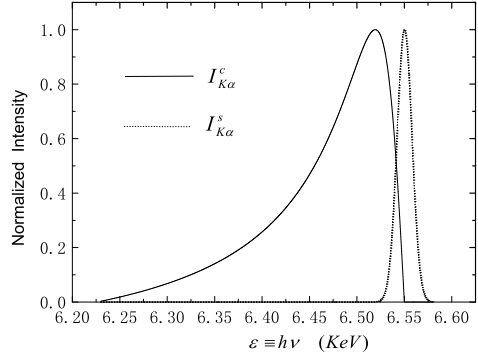}}
\caption{The calculated profile of Cerenkov line
$I_{K_{\alpha}}^{c}\sim \varepsilon$ of iron ion Fe$^{+21}$ in
optically thick case, where $\varepsilon \equiv h\nu$ is the
energy of line photon (You et al. 2003). The Cerenkov line-profile
is broad, asymmetric, and redshifted. The profile of a normal line
by spontaneous transition $I_{K_{\alpha}}^{s}\sim \varepsilon$ is
also plotted for the comparison.}
\end{figure}

\section{Cerenkov line mechanism}
We have proposed a new mechanism Cerenkov line emission for the
lines in X-ray afterglows. (Wang, Zhao \& You 2002). As we know,
Cerenkov radiation is produced when the particle velocity exceeds
the light speed in the medium. Furthermore, You \& Chen (1980)
argued that for relativistic electrons moving through a dense gas,
the Cerenkov effect will produce peculiar atomic/molecular
emission lines: Cerenkov lines. They also extended the work to the
X-ray band, such as Fe $K\alpha$ lines (You et al. 2000). Cerenkov
line emission has following remarkable features (see Fig. 1):
broad, asymmetrical, Cerenkov redshift and polarized if
relativistic electrons have an anisotropic velocity distribution.

Here, we emphasize the special importance of the Cerenkov redshift
which markedly strengthens the emergent intensity of the Cerenkov
emission line. For an optically thick dense gas, the emergent line
is determined both by the emission and absorption. The absorption
mechanisms for the normal line and Cerenkov line are extremely
different. The intensity of normal line, $I^n$ is greatly weakened
by the strong resonant absorption $k_{lu}(\nu_{lu})$ (the
subscripts $u$ and $l$ denote the upper and lower-levels) due to
the fact that the normal line is located exactly at the position
of the intrinsic frequency $\nu_{lu}$ where $k_{lu}$ is very
large. In the extreme case of a very dense gas, the emergent flux
has the continuum with a black body spectrum, and the normal line
disappeared, $I^n \sim 0$. However, the Cerenkov line, located at
$\nu<\nu_{lu}$ due to the redshift, can avoid the resonant
absorption because of $k_{lu}(\nu<\nu_{lu})\to 0$ (see the special
dispersion relations at a wavelength shown in Fig. 2). The main
absorption mechanism which affects the intensity of Cerenkov line
is the photoionization absorption $k_{bf}$ which is very small
compared with $k_{lu}$. Thus the Cerenkov line photons can easily
escape from deep inside a dense gas cloud, causing a strong
emergent line flux, only if the density of relativistic electrons
$N_e$ is high enough. In other words, the dense gas appears to be
more transparent for the Cerenkov line than normal line, which
makes it possible for the Cerenkov line mechanism to dominate over
the normal line mechanisms when the gas is very dense and there
exist abundant relativistic electrons in the emission region.

\begin{figure}
\centerline{\includegraphics[width=7cm]{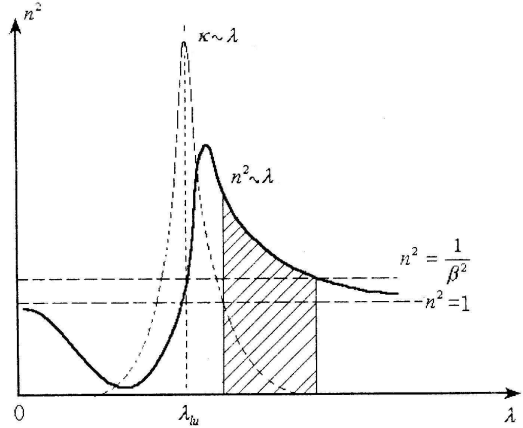}}
\caption{The relations between refractive index $n^2$ which is
proportional to emissivity and wavelength $\lambda$, and between
extinction coefficient $\kappa$ and $\lambda$. The Cerenkov
radiation survives in the shaded narrow region (You et al. 2000).}
\end{figure}

In the astrophysical processes of GRBs, a large amount of
relativistic electrons and dense gas regions could exist. For the
long burst afterglows localized so far, the host galaxies show
signs of star formation activity, where the high-density
environment is expected. Recent broadband observations of the
afterglows of GRB 000926 (Piro et al. 2001) and statistical
analysis (Reichart \& Price 2002) imply the evidence for a
fireball in a dense medium. When the ultra-relativistic blast
waves interact with the dense regions, very strong Cerenkov line
emission in X-ray band would be observed, and Cerenkov mechanism
may dominate in the case of the optically thick gases.
Importantly, broad line features are naturally explained in our
model (see Fig. 1).

The total intensity of Cerenkov line is (You et al. 2000) \beq
I^c=Y[{\rm ln}(1+X^2)+2(1-{\arctan X \over X})], \enq where
$Y\equiv {N_e\over 2}{C_1 \over k_{bf}}$, and $X\equiv
\sqrt{k_{bf} \over C_2}y_{lim}$, $N_e$ is the relativistic
electron density, $y_{lim}=C_0\gamma_0^2$, $\gamma_c$ is the
electron Lorentz factor, and \beqn
C_0&=&1.05\times10^{-76}\eps_{lu}^{-4}A_{ul}g_uN_{Fe}{S_l \over
g_l}, \nonumber \\
C_1&=&5.77\times10^{-53}\eps_{lu}^{-2}A_{ul}g_uN_{Fe}{S_l \over
g_l}, \nonumber \\
C_2&=&1.75\times10^{-87}\eps_{lu}^{-4}\Gamma_{lu}A_{ul}g_uN_{Fe}{S_l \over
g_l}, \nonumber \\
k_{bf}&=&8.4\times10^{-46}\eps_{lu}^{-3}N_{Fe}S_2. \enqn The
damping constant $\Gamma_{lu}$ is related to Einstein's
coefficient $A_{ul}$, $g_u$ is the statistical weights. For Fe
$K\alpha$ line, $\eps_{lu}=6.4$ keV, only levels $l=1, u=2$ are
considered, and $S_1=g_1=2, g_2=8$, and $S_2\sim5$ for the GRB
afterglow.

In the optically thick gas, we have found $X\gg1$, then we can
simplify the formulae of Cerenkov line intensity as $I^c\simeq
2Y({\rm ln}X-1)$ (Wang, Zhao \& You 2002). And we only take the
main contribution $2Y$ in our following estimation, so \beq
I^c\sim{C_1\over
k_{bf}}N_e\sim1.12\times10^{-7}\eps_{12}A_{21}N_e. \enq For the
optically thick case, the outward flux per unit area and per unit
time $\pi F \approx \pi I^c$. Assuming that there exist a lot of
spherical clouds with dense gas that is distributed homogeneously
in the circumburst environment, the cloud radius and number are
$R, N_c$ respectively, an isotropic fireball interacts with these
clouds. Therefore, the total line luminosity from the clouds is
$L^c=4\pi^2 R^2N_cI^c$. Defining a covering factor $C \equiv
{N_c\pi R^2/{4\pi D^2}}$, where $D$ is the distance between the
clouds and burst center, then \beq L^c=16\pi^2CD^2I^c\sim
8.2\times10^{32}C_{0.1}D_{16}^2N_e{\rm erg\ s^{-1}}, \enq where we
have taken $A_{21}=4.6\times10^{14}{\rm s^{-1}}$ and the
characteristic scales $C\sim 0.1, D \sim 10^{16}$cm. From Eq. (4),
we can clearly see that the total line luminosity by the Cerenkov
mechanism is strongly dependent on the electron density rather
than the iron abundance. Thus, we propose that some classes of GRB
models predict the presence of a very high relativistic electron
density, therefore the ultra-strong iron lines in the X-ray
afterglows can be explained probably without additional request
for initial and external conditions of iron-rich torus.

We also could estimate the relativistic electron density required
for X-ray lines from four GRB afterglows. Take the cosmological
model with $H_0=70{\rm km\ s^{-1}\ Mpc^{-1}}, \Omega_0=1/3,
\Omega_\Lambda=2/3$, and let $C_{0.1}=D_{16}=1$, we find that the
electron densities $N_e\sim 10^{10}-10^{11}{\rm cm^{-3}}$ are
similar to the one estimated by other models. We first simply
estimate the relativistic electron density from the fireball, $N_e
\sim E/{4\pi D^2\Delta_D \gamma m_pc^2}$, and taking the isotropic
energy $E\sim 10^{53}$erg, $\gamma \sim 100$, $\Delta_D \sim
\gamma cT$ (T is the duration $\sim 10$s), we obtained the density
$N_e\sim 10^8 {\rm cm^{-3}}$, which will be lower than our
requirement in the model. So we think the relativistic electrons
should come from other processes. We here propose that the
electrons would be produced by a pulsar wind from the central
millisecond magnetar (Thompson 1994) or $e^+ e^-$ outflow from the
Kerr black hole with magnetized torus (MacFadyen \& Woosley 1999)
as the delayed injection after GRB events. Our interpretation is
similar to that of Rees \& M\'esz\'aros (2000), but we consider
the $e^+ e^-$ outflow instead of electromagnetic flux. The
luminosity is required as high as $L\sim 4\pi D^2N_e c\gamma_e m_e
c^2 \sim 10^{48} {\rm erg\ s^{-1}}$, where we take $N_e \sim
10^{10}{\rm cm^{-3}}, \gamma_e \sim 10$ at the distance of
$10^{16}$cm. The luminosity could be satisfied by the central
compact objects with the acceptable parameters such as very strong
magnetic field. Because the scattering cross section of
relativistic electrons will be near to the zero due to the
Klein-Nishina formula at the very high frequency(due to the
Doppler effect), we need not worry about the electron scattering
process will greatly affect the line profile and intensity.

Furthermore, we also have considered the optically thin case for
the Cerenkov line emission mechanism (Wang 2003). In this case,
The emission intensity will be dependent on both the electron
density and iron abundance (in a form $\propto X^2Y$). Meanwhile,
the computations show that only a small mass of iron about
$10^{-6}M_\odot$ is needed to produce the high line luminosity
within the emission region of $\sim 10^{16}$cm, so a normal
homogeneous medium around the GRB progenitor could meet the
requirement.

Finally, the research and development of thirty years on gamma ray
bursts more and more make the astronomers believe that there
probably exists a unified picture for gamma ray bursts. Recently,
some researchers proposed the jet unified models of short and long
gamma-ray bursts, X-ray rich gamma-ray bursts, and X-ray flashes
(Lamb et al. 2004; Yamazaki et al. 2004). The jet models assume
the different classes of gamma-ray bursts have a standard total
energy within a range of $10^{50}-10^{51}$ erg (Frail et al.
2001). However, the standard energy is much lower than the total
energy limit determined from the X-ray emission lines, $\sim
10^{52}$ erg (Ghisellini et al. 2002). Since based on the
fluorescence or recombination, the line radiation is isotropic,
and the burst energy cannot be lower than the X-ray afterglow
energy, so the inconsistency still exists. But the Cerenkov line
mechanism has not the problem and could make the jet unified
picture more convincible, because the Cerenkov line intensity can
be independent of the prompt energy of gamma-ray bursts, and if
the relativistic electrons distribute not in the homogeneous
direction, the Cerenkov line emissions are not isotropic too, the
``beaming'' effect will decreases the total X-ray afterglow energy
to the same order of the standard energy reservoir.

\begin{table}
\scriptsize \caption{The properties of X-ray emission lines in
GRBs. Here the cosmological model $(\Omega_M, \Omega_\Lambda,
h)=(1/3,2/3,0.7)$ is taken, the redshift of GRB 000214 is
determined only based on Fe $K\alpha$ emission identification. Two
lines is discovered in GRB 991216, generally, one is identified as
the fluorescent line, the other as the recombination.}
\begin{center}
\begin{tabular}{l c c c c c c l}
\hline
GRB & Detector & z & Line Energy & Width & Intensity & Line Luminosity & Reference\\
\  & \ & \ & keV & keV & 10$^{-5}$ ph cm$^{-2}$s$^{-1}$ & erg s$^{-1}$ & \  \\
\hline
970508 & BeppoSAX & 0.835 & 3.4 & $\leq 0.5$ & 5.0 & 2.7$\times 10^{44}$ & 1,2 \\
970828 & ASCA & 0.957 & 5.04 & 0.31 & 1.9 & 1.1$\times 10^{44}$ & 3,4  \\
991216 & Chandra & 1.02 & 3.49 & 0.23 & 3.2 & 4.0$\times 10^{44}$ & 5,6  \\
\  & \ & \  & 4.4 & 1.0 & 3.8  & 4.8$\times 10^{44}$ & 5  \\
000214 & BeppoSAX & 0.47 & 4.7 & 0.2 & 0.9 & 4.0$\times 10^{43}$ & 7 \\
\hline
\end{tabular}
References: 1. Piro et al. 1999; 2. Metzger et al. 1997; 3.
Yoshida et al. 1999; 4. Djorgovski et al. 2001; 5. Piro et al.
2000; 6. Vreeswijk et al. 2000; 7. Antonelli et al. 2000.
\end{center}
\end{table}

\end{document}